# Diagnosable-by-Design Model-Driven Development for IEC 61499 Industrial Cyber-Physical Systems

Barry Dowdeswell, Roopak Sinha, and Stephen G. MacDonell
*Department of IT & Software Engineering*
*School of Engineering, Computing and Mathematical Sciences*
Auckland University of Technology, Auckland, New Zealand
barry.dowdeswell@aut.ac.nz, roopak.sinha@aut.ac.nz, stephen.macdonell@aut.ac.nz

**Abstract**
*Integrating the design and creation of fault identification and diagnostic capabilities into Model-Driven Development methodologies is one approach to enhancing the resilience of Industrial Cyber-Physical Systems. We present a Fault Diagnostic Engine designed to recognise and diagnose faults in IEC 61499 Function Block Applications. Using diagnostic agents that interact directly with the target application, we demonstrate fault monitoring and analysis techniques and as well as failure scenario intervention. By designing and building fault diagnostic resources during early phases of Model-Driven Development, both iterative testing and long-term fault management capabilities can be created. While applying and refining appropriate model artifacts, we demonstrate that the concurrent development of function blocks alongside fault management capabilities is both feasible and worthwhile.*

## 1. INTRODUCTION

Industrial Cyber-Physical Systems (ICPS) built upon the IEC 61499 Function Block Architecture [1] are mechanisms that blend computing elements with sensors and actuators. By forming networks of distributed, interconnected devices, they interact with the physical environments they are deployed in to control machinery.

Testing ICPS during their design and development phases is a well-established engineering discipline [2]. However, many of the faults these devices later exhibit do not occur during these early phases. The long-term interactions that ICPS have with their physical environments can expose unexpected faults that are far more problematic and intriguing. Design methodologies should address the creation of long-term fault management capabilities, enabling ICPS to cope with failures both within the hardware they rely on and the control algorithms that guide their operations.

ICPS interact with their environments through physical hardware. Timely responses to control mechanical apparatus such as aircraft ailerons and landing gear are crucial to the safe operation of these systems. However, it is not enough to consider the computational and electromechanical elements of an ICPS separately [3]. Rather, it is at the *intersection* of the cyber and the physical that the most challenging fault scenarios emerge.

This paper presents a Fault Diagnostic Engine (FDE) designed to identify and diagnose faults in IEC 61499 Function Block Applications (FBAs). The FDE monitors system-wide behavior using appropriate fault detection strategies, watching to see if the control commands issued result in the correct operation of the ICPS. When the symptoms of a fault are observed, appropriate diagnostic tests can be applied to identify, classify and isolate the faulty ICPS components.

Diagnostic agents with domain knowledge of the IEC 61499 architecture that are operating under the FDE interact directly with Function Blocks (FBs). We demonstrate how to implement custom Service Interface Function Blocks (SIFBs) wired dynamically into the application running under the FORTE runtime [4]. These SIFBs interact with teams of diagnostic agents operating within the GORITE Multi-Agent System [5]. Techniques for both long-term observation and fault moni-toring as well as failure scenario intervention are presented. By designing and developing fault diagnostic resources during the Model-Driven Development (MDD) cycles, both iterative testing and long-term fault management capabilities can be created to support engineers throughout the life cycle of the ICPS.

This paper demonstrates the modeling of the FBA while developing diagnostic capabilities in-parallel. It builds on our earlier scoping survey of fault diagnostic methodologies [6]. We detail the creation of diagnosable-by-design artifacts using a combination of SysML and IEC 61499 diagrams that provide diagnostic capabilities to accompany the FBA through its entire life-cycle beyond its design stages.

Section 2 outlines IEC 61499 and the event-driven control this architecture provides. Section 3 presents a Heating, Ventilation and Air-Conditioning (HVAC) co-simulation that was created to support the development of the FBs used in the ICPS. The requirements elucidation that preceded the modeling phases illustrate how MDD approaches encourage diagnosable-by-design principles that identify potential fault pathways. We also describe the type of faults that might be encountered in IEC 61499 systems such as these. Section 4 examines the operation of the diagnostic agents in more detail, evaluating the performance of the FDE when operating with multiple, simultaneous agent instances. In Section 5 we propose future directions for this research.

## 2. BACKGROUND TO IEC 61499 BASED ICPS

ICPS built with IEC 61499 function blocks rely on sensors to capture information about the physical tasks they are man-aging. Electromechanical actuators such as valves or motors are then used to perform physical tasks in their environment. The control of sensors, actuators and any control and data transformations are carried out through software elements called *function blocks* (FBs).

IEC 61499 function blocks provide an object-oriented software architecture for building ICPS [7], [8]. Mediated by a set of IEC standards, FBs address many of the shortcomings of earlier embedded controller design approaches [1]. The architecture encourages the flexible reuse of components as well as reliable data exchange between distributed sub-applications by employing secure, industrial-strength connectivity [10].

Each FB facilitates one or more of the requirements of the ICPS control functionality. Fig. 1 shows a FB that forms part of the room controller for the HVAC system. F_TO_C_CONV is responsible for converting Fahrenheit temperatures to Celsius.

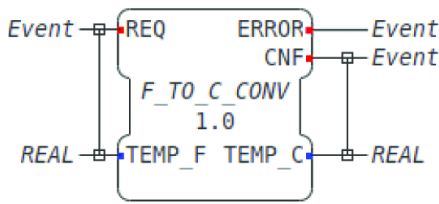

Fig. 1: Function block created for the Room controller.

This FB receives new temperature readings to convert by responding to an *Input Event* received on the port REQ, sent from another FB it is connected to. This event triggers the acquisition of a new Fahrenheit temperature value via the *Input Data* port TEMP_F. Algorithms within the FB then perform the temperature conversion. If successful, the FB passes the Celsius temperature out via the *Output Data* port TEMP_C before triggering the *Output Event* CNF. If the temperature cannot be converted, the FB signals an error by triggering the event ERROR. This event-driven control is typical of ICPS built with IEC 61499.

## 3. MODEL-DRIVEN DEVELOPMENT WITH DIAGNOSTICS

Most design and development tasks for ICPS are complex optimization problems for real-time systems [11]. The physical characteristics of sensors and actuators add a level of complexity to both the functional and quality requirements [12], [13]. ICPS often operate under strict timing constraints where FBs have to interact with real-world events that are both asynchronous and exhibit parallelism [14]. The operations they perform in their environment demand timely responses and often, high-precision [11]. Hence, designers must remain cognizant of the physical characteristics of the devices they are interfacing to.

Pre-Requirements statements are the starting point from which designers begin to iteratively explore design options through rounds of engineering analysis, interpretation and refinement. Pre-Requirements are usually written in free-form natural language, are often incomplete, and are sometimes ambiguous [12]. They describe what is needed without considering how the ICPS might meet those needs [15]. However, diagnostic considerations are seldom considered during this phase. The early ideas for the room controller included the concept sketch shown in Fig. 2. The sketch mentions the quality requirements of sub-second response times for button presses and display updates.

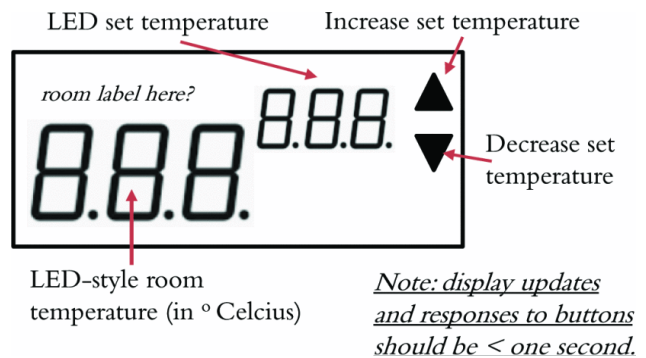

Fig. 2: Early-stage concept sketch of the room controller.

Table I illustrates some of the refined requirements. These rely on an ontology that describes the proposed room controller elements and their attributes. Ambiguous terms from the Pre-Requirements are replaced by well-defined terms such as Temperature and HVAC_Controller to identify elements as well as operations such as decreases and notify.

ISO Standard 25010 defines quality attributes such as performance and timeliness [16]. Derived from the ISO Software Engineering Product Quality standard ISO/IEC 9126, ISO 25010 focuses on software-intensive systems such as ICPS. Table I shows the relationship between the operational requirements and the corresponding quality attributes that the design must satisfy for each requirement.

### A. Modeling with SysML diagrams and 4diac

The SysML Block Diagram shown in Fig. 3 identifies broad functional units needed to meet the requirements.

There is a clear distinction between the way the Pre-Requirements explain to the designers what is needed and the detailed designs that emerge after the requirement specification has been written.

TABLE I: Example Functional and Quality Requirements

| Requirement | Description | ISO 25010 Quality Attribute |
|---|---|---|
| RQ 1.0 | when Temperature.increases or Temperature.decreases | |
| NFR 1.1 | HVAC_Controller.notify occurs in < 500 ms. | 4.2.2 Performance Efficiency |
| NFR 1.2 | DisplayedTemperature.update occurs in < 250 ms | 4.2.2.1 Time behavior |
| RQ 3.0 | when SetTemperatureDown.pressed then | |
| NFR 3.1 | HVAC_Controller.notify occurs in < 500 ms. | 4.2.2 Performance Efficiency |
| NFR 3.2 | SetTemperature.update occurs in < 100 ms | 4.2.2.1 Time behavior |

Each block on the SysML Block Diagram is realized by one or more function blocks (FBs). Once appropriate sensors and actuators are considered, design constraints emerge as the requirements are refined. It is during this phase of the modeling that an understanding of the diagnostic needs begins to emerge. For example, the Pre-Requirements stated that all temperatures should be displayed in Celsius. However, the temperature probe that best fits the cost and technical needs only provides readings in Fahrenheit. This demanded the addition of F_TO_C_CONV, which was also tasked with performing error checks on the temperature probe readings. This sensor was also noted as a potential point-of-failure.

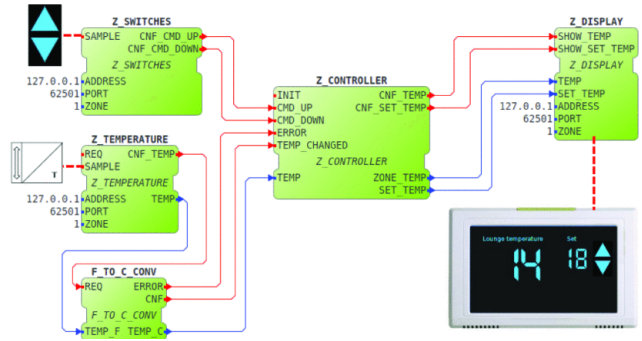

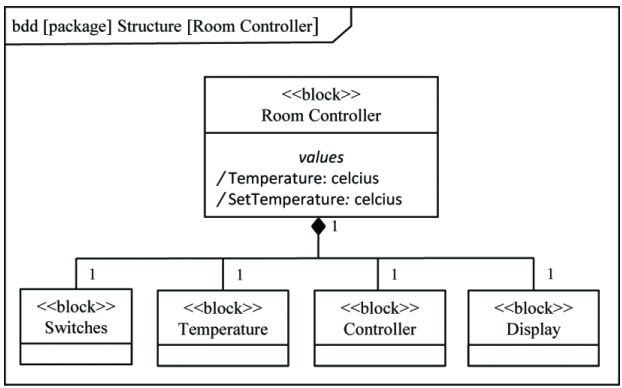

Fig. 3: SysML Block Diagram for the function blocks.

Fig. 4 shows the next design artifact after the application has been modeled in the 4diac Function Block Development System [17]. Each FB is a class instance created from a primitive IEC 61499 FB type. The IEC 61499 convention places all public in-bound events and data inputs on the left of each FB symbol while all outbound events and data outputs appear on the right. Connections between sub-sections further delineate the separations of concerns, modeling the flow of data between elements. This diagram therefore presents information in a way more appropriate to the FB architecture than a SysML Class Diagram would. It also captures the thinking of the designer in a less abstract way that is closer to what the final application will look like.

While designing applications in this way, there is a subtle but natural inclination to drive the flow of information in the diagram from left to right. This is in line with the IEC 61499 convention. The temperature capture and conversion sub-section felt natural to place on the left. Similarly, it was logical to position the Z_CONTROLLER that co-ordinates the information from all sensors and actuators in the center of the model. While the display and on-screen switches will later be realized as a single integrated unit, their functions were separated in the model with the switches expressed as inputs on the left and the display as outputs on the right. Creating such uniformity of expression is a key goal of SysML [18]. Well-understood diagramming conventions not only speed up the design process, they remove ambiguity, foster better communication between stakeholders, and lead to specifications that can be refined iteratively [14]. As we show later in Section 3-B, this convention also makes it easier to identify fault pathways.

Fig. 4: Room Controller implemented with Function Blocks.

Each FB operates its own internal state machine called an *Execution Control Chart* (ECC) [19]. Fig. 5 shows some of the event sequences and data exchange between FBs in the room controller and the HVAC main controller.

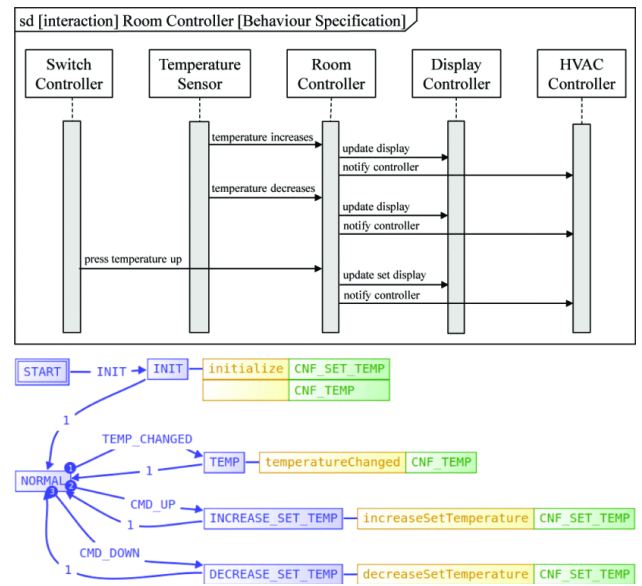

Fig. 5: Partial SysML Sequence Diagram with the individual Z_CONTROLLER Execution Control Chart (ECC).

The Z_CONTROLLER is constructed from a fundamental IEC 61499 template called a Basic Function Block (BFB). State changes in the ECC trigger input and output events that, with their accompanying data, exchange control information between the individual FBs in the FBA. SysML Sequence diagrams document high-level interactions while individual ECCs, modeled in 4diac, detail state transitions within each FB in greater detail. For each state transition shown on the ECC, one or more algorithms can be triggered to process data and make decisions about outputs. Algorithms may also cause subsequent state changes.

### B. Modeling faults and creating diagnosis strategies

Any change in the operation of an ICPS that leads to degraded performance or unacceptable behavior is defined as a *fault* [20]. Recognizing the symptoms of a fault requires knowing the difference between what is normal behavior and misbehavior by one or more components of the ICPS [21]. Determining a baseline of what constitutes normal behavior is key to modeling successful ICPS fault management strategies. The systematic diagnosis performed after a problem is detected then seeks to isolate and identify exactly which device or sub-system is causing the errant behavior. Complicating this, the presence of multiple, simultaneous faults in large systems with a myriad of sensors, actuators, and control elements demands innovative approaches [22], [23].

**Sensor faults.** The Z_TEMPERATURE SIFB shown in Fig. 4 is connected to a Platinum Resistive Thermometer (PTR). PTRs are sensors that change their electrical resistance linearly as the temperature changes. Properly calibrated PTRs, fitted with the appropriate electronics, are more accurate and stable than other types of temperature sensors. However, PTRs are fragile and exhibit a number of distinct failure modes [24]. The resistive element can crack when stressed, causing it to have a resistance that is out-of-tolerance with its calibrated value. A failure of the electronic components in the sensors interface can cause the internal current and voltage generator to misreport the temperature. Intermittent connections also cause the sensor to deliver readings that vary widely each time the FB samples the sensor.

**Actuator electrical and mechanical faults.** Not all devices in an ICPS have the ability to report internal failures. Other types of faults in the HVAC heating and cooling machinery result in the ICPS failing to control its environment properly. These include problems with actuators that control the airflow in ducting. In other situations, sensors that misreport the position of a control vent can result in the icing-up of a duct, blocking airflows. These types of faults result in the ICPS believing that the correct amount of heating or cooling is being delivered to a room when it is not. Without appropriate diagnostic monitoring, such faults may not always be detected by the ICPS itself.

**Software faults.** Software algorithms that implement the cyber parts of a ICPS can also fail, often as a result of updates applied after an earlier successful commissioning of a system. Software misbehavior is sometimes indistinguishable from sensor or actuator failures. During fault identification, the FDE applies an iterative divide-and-conquer approach that first isolates the sensor or other electromechanical hardware from the FBs in the section of the ICPS that is being examined. By exercising each component separately with pre-defined diagnostic plans, the agents determine which parts are still acting normally and which components are exhibiting fault signatures.

### C. Creating fault scenarios from models

Each FB is a class, but it is not practical to create typical Test-Driven Design (TDD) Unit Tests for them. The 4diac IDE provides a way of manually exercising the inputs and outputs of a single FB. However, automatically iterating a range of test values in 4diac is not yet supported. Therefore, exercising multiple FBs in the IDE to diagnose faults with data sets is not possible. Hametner et al. [25] propose one approach to TDD after first identifying what the fundamental unit of an FBA is that a unit test could be applied to. FBs meet the criteria of being software components [26], [27] since they encapsulate both their ECCs and their internal data. Similarly, Christensen [28] proposes a Model-View-Controller with Diagnostics (MVCD) design pattern that incorporates an early-stage diagnostics design approach. Hence testing approaches usually approach FBs as black-box components with observable behaviors rather than attempting to directly unit test them as functions in classes [29].

The proposed MDD with Diagnostics approach addresses these difficulties by identifying potential faults and fault pathways in FBs concurrently while they are being designed. Diagnostic tests are coded to provide appropriate test coverage by capturing related data values from one or more FBs while they are being exercised. These can be executed individually by the FDE agents on-demand. As the design of each FB progresses incrementally, potential fault scenarios are identified iteratively, making the diagnostic test suite progressively more comprehensive. FBs are designed to be re-usable, so subsets of fault tests and test values are accumulated to form a library of resources for subsequent development.

Creating fault identification tests early improves both the intrinsic design and the reliability of each FB. As parts of the FBA are completed, groups of related FBs can be exercised to help refine the design for the next stages. Fig. 6 maps the fault pathways through the temperature sensor sub-system of the FBA.

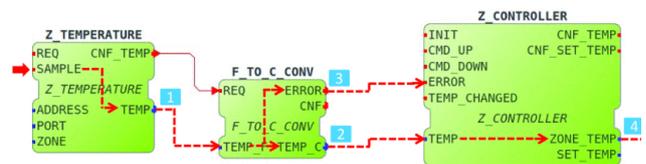

Fig. 6: Temperature sensor sub-system fault pathways.

The Z_TEMPERATURE SAMPLE event is triggered when the FB has acquired a value from the temperature sensor. The first accessible Diagnostic Point (DP) on this FB is the output TEMP, which passes the reading into F_TO_C_CONV TEMP_F. This FB also checks the range of reasonable temperatures. In-range readings pass out via TEMP_C to TEMP on Z_CONTROLLER. Irregular readings that deviate significantly from the last reading are considered outliers. If this behavior persists, the ERROR output will fire, indicating to Z_CONTROLLER that unexpected readings have been detected. The blue squares identify DPs along this fault pathway that can be monitored.

## 4. AGENT DIAGNOSTICS

Once all the potential fault pathways have been identified, the FDE is able to deploy a harness to capture data from designated diagnostic points. Fig. 7 shows an AGENT_GATE function block that the FDE *rewire()* command has inserted automatically into the FBA to capture activity at this Diagnostic Point (DP). The FDE provides multiple FIFO input and output packet queues to ensure that the diagnostic scripts can work asynchronously without losing track of the telemetry exchanged with the FBA.

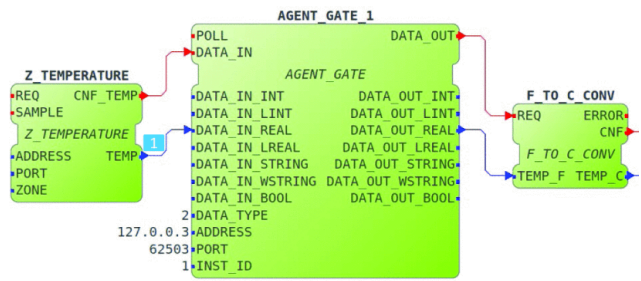

Fig. 7: Agent function block wired into a diagnostic point.

While the rewired FBA is operating normally, the AGENT_GATE instances pass events and data transparently through themselves. They also send telemetry back to the FDE agents, identifying themselves in the transmission packet so that the FDE can correlate readings from multiple diagnostic points simultaneously. This *Monitor* mode is read-only.

The simbIoTe simulator also provides an HMI that allows button presses to be generated while emulating the temperature displays for the FBA during development [30]. Within sim-bIoTe, the Environment component generates a range of temperatures that vary stochastically, driven by a simple algorithm. This simulation approach allows rapid prototyping of the FBA while supporting co-simulation as hardware components are completed, replacing parts of simulated hardware.

In Monitor mode, the FDE agents track data passing through the DPs, watching values along the potential fault pathways identified earlier. Once an outlier is detected, or if a signal being followed does not emerge from a pathway within an expected interval, the FDE can intervene. The agents can then intervene and execute diagnostic plans by instructing the AGENT_GATE instances to ring-fence parts of the FBS so it can be diagnosed. The temperature sensor sub-system between DPs shown in Fig. 6 is isolated by issuing a *gateClose* command to the AGENT_GATE instances at DP 1 and DP 5. This blocks data coming in from the temperature sensor and going out of the Z_CONTROLLER. A pre-defined set of diagnostic test values is then sent through the fault path by injecting values into DP 1 and reconciling the outputs captured at the other DPs. Fahrenheit diagnostic values injected into DP 1 should generate matching Celsius values at DP 2. During a diagnostic run a temperature of absolute zero resulted in F_TO_C_CONV triggering its ERROR output correctly at all times. As expected, the Z_CONTROLLER ZONE_TEMP value did not appear during this error condition.

This divide-and-conquer approach allows the FDE agent to update its beliefs about what may be wrong. For example, if all the diagnostic values injected into the fault pathways result in correct values emerging through the other diagnostic points, the agent concludes that the temperature sensor is faulty. Conversely, if the FDE *compare*() function cannot correlate temperatures being processed by F_TO_C_CONV, then it concludes that this FB has a faulty algorithm. IEC 61499 applications are often partitioned into distributed *sub-applications* so they can be run on separate computing devices. The complete HVAC is built up from multiple room controllers connected to a HVAC Main Controller FBA running on its own hardware. The FDE supports teams of co-operating agents who are deployed to watch and diagnose sub-applications. Using back-channel communications built into the FDE, agents exchange beliefs about what is happening in their sub-applications to build up a shared understanding of how the system is behaving.

Distributed GORITE agents were used to investigate the blocked duct scenario outlined in Section III-B. While the FDE can operate locally, we chose to deploy it on a separate system during all tests. This allowed us to investigate more aggressive fault scenarios that involved complete failures of specific sub-applications. In Monitor mode, the agents applied a Bayesian analysis of the temperatures being read by the zone. Once a temperature change is requested by a user, the HVAC machinery should start a heating or cooling cycle. Based on the ASHRAE Standard 55 [31], we estimated a suitable rate of change for occupant thermal comfort of no more than 0.3°C/min. When simbIoTe simulated a rate that was significantly different, the agents correctly launched a diagnostic intervention, working through subsystems to identify the fault location. We also simulated the situation where there was no change in response to an occupants request which generated a similar response by the FDE.

### A. Results from monitoring and diagnostic interventions

The FDE was evaluated with ten AGENT_GATE instances on different DPs, returning data in Monitor mode. The FBA samples its switches and sensors at 100ms intervals to meet the quality requirements. No discernible degradation in performance was evident. When an agent decided to intervene and switch to *Diagnostic* mode, it issues as *gate*() command to the relevent AGENT_GATE instances within that sub-system. Exercising the diagnostic test in this mode while other parts of the FBA were still operating showed no discernible performance issues. This can be attributed to the AGENT_GATE being coded directly in C++ and cycling on an interval timer that is configured dynamically by the *rewire*() command to sample more frequently than the fastest FB identified in the FBA.

A number of software faults were simulated by introducing a random error in the Fahrenheit to Celsius conversion for specific temperature ranges. While in *Monitor* mode, the FDE agent detected the differing outputs at DP 1 and 2 while recognizing that the ERROR event at DP 3 had not occurred. After intervening in *Diagnostic* mode, it proposed that there was a fault in `F_TO_C_CONV`.

## 5. CONCLUSIONS AND FUTURE WORK

Applying a diagnosable-by-design, model-driven development approach proposed in this paper is one way of building resilience into an ICPS. Future directions for this work include a level of compliance with the syntax of testing languages used in products such as Selenium [32]. Safety-critical aspects also need to be considered further: isolating critical sub-systems after first securing what is being controlled is a vital aspect of fault management. Building such safeguards into diagnostic library resources remains a fascinating area to examine further.

Techniques such as these have deliver practical, early-stage productivity gains to engineers if they are to see wider adoption. This approach is one way of addressing the limitations that IEC 61499 places on early stage testing while providing a foundation for long-term diagnosability and resilience of ICPS.

## REFERENCES

[1] I. E. Commission et al., "Function blocks–Part 1: Architecture," International Electrotechnical Commission, Geneva, Switzerland, Tech. Rep. IEC, 2013.
[2] X. Zhou, X. Gou, T. Huang, and S. Yang, "Review on Testing of Cyber Physical Systems: Methods and Testbeds," IEEE Access, vol. 6, pp. 52 179–52 194, 2018.
[3] E. A. Lee and S. A. Seshia, Introduction to embedded systems: A cyber-physical systems approach. MIT Press, 2016.
[4] 4DIAC-RTE (FORTE): IEC 61499 Compliant Runtime Environment, 2019. [Online]. Available: https://www.eclipse.org/4diac/
[5] D. Jarvis, J. Jarvis, R. Rönnquist, and L. C. Jain, "Multi-Agent Systems," in Multiagent Systems and Applications. Springer, 2013, pp. 1–12.
[6] B. Dowdeswell, R. Sinha, and S. G. MacDonell, "Finding faults: A scoping study of fault diagnostics for Industrial Cyber-Physical Systems," Journal of Systems and Software, vol. 168, p. 110638, 2020.
[7] L. H. Yoong, P. S. Roop, Z. E. Bhatti, and M. M. Kuo, "Formal Model for IEC 61499 Function Blocks," in Model-Driven Design Using IEC 61499. Springer, 2015, pp. 65–91.
[8] W. Dai, V. Vyatkin, and J. Christensen, "Applying IEC 61499 de- sign paradigms: Object-oriented programming, component-based design, and service-oriented architecture," in Distributed Control Applications: Guidelines, Design Patterns, and Application Examples with the IEC 61499. CRC Press, 2015.
[9] I. E. Commission et al., "Function blocks–Part2: Software Tool Requirements," International Electrotechnical Commission, Geneva, Switzerland, Tech. Rep. IEC, 2013.
[10] A. Tanveer, R. Sinha, and S. G. MacDonell, "On Design-time Security in IEC 61499 Systems: Conceptualisation, Implementation, and Feasibility," in 2018 IEEE 16th International Conference on Industrial Informatics (INDIN). IEEE, 2018, pp. 778–785.
[11] P. Zave, "An operational approach to requirements specification for embedded systems," IEEE transactions on Software Engineering, no. 3, pp. 250–269, 1982.
[12] B. Vogel-Heuser, D. Schütz, T. Frank, and C. Legat, "Model-driven engineering of manufacturing automation software projects–A SysML- based approach," Mechatronics, vol. 24, no. 7, pp. 883–897, 2014.
[13] A. A. Cabrera, M. Foeken, O. Tekin, K. Woestenenk, M. Erden, B. De Schutter, M. Van Tooren, R. Babuška, F. J. van Houten, and T. Tomiyama, "Towards automation of control software: A review of challenges in mechatronic design," Mechatronics, vol. 20, no. 8, pp. 876–886, 2010.
[14] V. Vyatkin, "IEC 61499 as enabler of distributed and intelligent automation: State-of-the-art review," IEEE transactions on Industrial Informatics, vol. 7, no. 4, pp. 768–781, 2011.
[15] C.-W. Yang, V. Dubinin, and V. Vyatkin, "Automatic Generation of Control Flow from Requirements for Distributed Smart Grid Automation Control," IEEE Transactions on Industrial Informatics, 2019.
[16] O. I. de Normalisation, Systems and Software Engineering: Systems and Software Quality Requirements and Evaluation (SQuaRE): System and Software Quality Models. ISO/IEC, 2011.
[17] T. Strasser, M. Rooker, G. Ebenhofer, A. Zoitl, C. Sünder, A. Valentini, and A. Martel, "Framework for distributed industrial automation and control (4DIAC)," in Industrial Informatics, 2008. INDIN 2008. 6th IEEE International Conference on. IEEE, 2008, pp. 283–288.
[18] A. A. Shah, A. A. Kerzhner, D. Schaefer, and C. J. Paredis, "Multi-view modeling to support embedded systems engineering in SysML," in Graph transformations and model-driven engineering. Springer, 2010, pp. 580–601.
[19] P. Lindgren, M. Lindner, D. Pereira, and L. M. Pinho, "A Formal Perspective on IEC 61499 Execution Control Chart Semantics," in Trustcom/BigDataSE/ISPA, 2015 IEEE, vol. 3. IEEE, 2015, pp. 293–300.
[20] T. R. Thombare and L. Dole, "Review on fault diagnosis model in automobile," in Computational Intelligence and Computing Research (ICCIC), 2014 IEEE International Conference on. IEEE, 2014, pp. 1–4.
[21] F. Harirchi and N. Ozay, "Guaranteed model-based fault detection in cyber-physical systems: A model invalidation approach," arXiv, 2016.
[22] E. Bartocci, N. Manjunath, L. Mariani, C. Mateis, and D. Ničković, "Automatic Failure Explanation in CPS Models," in International Conference on Software Engineering and Formal Methods. Springer, 2019, pp. 69–86.
[23] J. De Kleer and B. C. Williams, "Diagnosing multiple faults," Artificial intelligence, vol. 32, no. 1, pp. 97–130, 1987.
[24] Beamex, "Resistance measurement; 2,3 and 4 wire connection," 2020. [Online]. Available: https://blog.beamex.com/ resistance-measurement- 2- 3- or- 4- wire-connection
[25] R. Hametner, I. Hegny, and A. Zoitl, "A unit-test framework for event- driven control components modeled in IEC 61499," in Proceedings of the 2014 IEEE Emerging Technology and Factory Automation (ETFA). IEEE, 2014, pp. 1–8.
[26] C. Sunder, A. Zoitll, J. H. Christensen, H. Steininger, and J. Rritsche, "Considering iec 61131-3 and iec 61499 in the context of component frameworks," in 2008 6th IEEE International Conference on Industrial Informatics. IEEE, 2008, pp. 277–282.
[27] D. Winkler, R. Hametner, and S. Biffl, "Automation component aspects for efficient unit testing," in 2009 IEEE Conference on Emerging Technologies & Factory Automation. IEEE, 2009, pp. 1–8.
[28] J. H. Christensen, "Design Patterns, Frameworks, and Methodologies," Distributed Control Applications: Guidelines, Design Patterns, and Application Examples with the IEC 61499, p. 27, 2017.
[29] T. Hussain and G. Frey, "UML-based development process for IEC 61499 with automatic test-case generation," in 2006 IEEE Conference on Emerging Technologies and Factory Automation. IEEE, 2006, pp. 1277–1284.
[30] The simbIoTe Physical Simulation Environment for IoT devices., 2020. [Online]. Available: https://github.com/badger-dowdeswell/simbIoTe.git
[31] A. Standard, "ASHRAE Standard 55-2004," Thermal environmental conditions for human occupancy, vol. 3, 2004.
[32] The Selenium Testing Environment., 2020. [Online]. Available: https://www.selenium.dev/